\documentstyle[preprint,aps]{revtex}
\begin{document}
\draft
\preprint{\hbox to \hsize{\hfil\vtop{\hbox{IASSNS-HEP-99/36}
\hbox{April, 1999}}}}

\title{State Vector Collapse Probabilities and Separability\\
of Independent Systems in Hughston's\\
Stochastic Extension of the Schr\"odinger Equation \\
}
\author{Stephen L. Adler\\}
\address{Institute for Advanced Study\\
Princeton, NJ 08540\\
}
\author{Lawrence P. Horwitz
\footnote{On leave from School of Physics and Astronomy, 
Raymond and Beverly Sackler Faculty 
of Exact Sciences, Tel Aviv University, Ramat Aviv, Israel, and 
Department of Physics, Bar Ilan University, Ramat Gan, Israel.}}
\address{Institute for Advanced Study\\
Princeton, NJ 08540\\
}

\maketitle

\leftline{Send correspondence to:}

{\leftline{Stephen L. Adler}
\leftline{Institute for Advanced Study}
\leftline{Olden Lane, Princeton, NJ 08540}
\leftline{Phone 609-734-8051; FAX 609-924-8399; email adler@ias.edu}}

\begin{abstract}
We give a general proof that Hughston's stochastic extension of the 
Schr\"odinger equation leads to state vector collapse to energy eigenstates, 
with collapse probabilities given by the quantum mechanical probabilities 
computed from the initial state.  We also show that for a system 
composed of independent subsystems, Hughston's equation separates into 
similar independent equations for the each of the subsystems, correlated  
only through the common Wiener process that drives the state reduction. 
\end{abstract}                          

A substantial body of work [1] has addressed the problem of state vector 
collapse by proposing that the Schr\"odinger equation be modified to 
include a stochastic process, presumably arising from physics at a deeper 
level, that drives the collapse  process.  Although interesting models 
have been constructed, there so far has been no demonstration that for a 
generic Hamiltonian, one can construct a stochastic dynamics that collapses  
the state vector with the correct quantum mechanical probabilities.  
Part of the problem has been that most earlier work has used stochastic  
equations that do not preserve state vector normalization, requiring 
additional ad hoc assumptions to give a consistent 
physical interpretation.  

Various authors [2] have proposed rewriting the Schr\"odinger 
equation as an equivalent dynamics on projective Hilbert space, i.e., on 
the space of rays, a formulation in which the imposition of a state vector 
normalization condition is not needed.  Within this framework, Hughston [3] 
has proposed a simple stochastic extension of the Schr\"odinger equation, 
constructed solely from the Hamiltonian function, and has shown that his 
equation leads to state vector reduction to an energy eigenstate, with   
energy conservation in the mean throughout the reduction process.    
In the simplest spin-1/2 case, Hughston exhibits an explicit solution 
that shows that his equation leads to collapse with the correct quantum 
mechanical probabilities, but the issue of collapse probabilities in the 
general case has remained open.   In this Letter, we shall give a general 
proof that Hughston's equation leads to state vector collapse to energy 
eigenstates with the 
correct quantum mechanical probabilities, using the martingale or ``gambler's 
ruin'' argument pioneered by Pearle [4].  We shall also show that Hughston's 
equation separates into independent equations of similar structure for a 
wave function constructed as the product of independent subsystem wave 
functions.  

We begin by explaining the basic elements needed to understand Hughston's 
equation, working in an $n+1$ dimensional Hilbert space.  We denote the 
general state vector in this space by $| z \rangle$, with $z$ a shorthand 
for the complex projections $z^1,z^2,...,z^{n+1}$ of the state vector on an 
arbitrary fixed basis.  Letting  $F$ be an arbitrary Hermitian operator, and 
using the summation convention that repeated indices are summed over their 
range, we define 

\begin{mathletters}
\label{allequations}
\begin{equation}
(F) \equiv { \langle z | F | z \rangle   \over \langle z |z \rangle } 
= { \overline z^{\alpha} F_{\alpha \beta} z^{\beta} \over  
\overline z^{\gamma} z^{\gamma} }~~~,
\label{equationa}
\end{equation}
so that $(F)$ is the expectation of the operator $F$ 
in the state $|z\rangle$, 
independent of the ray representative and normalization 
chosen for this state.  
Note that in this notation $(F^2)$ and $(F)^2$ are not the same; their 
difference is in fact the variance $[\Delta F]^2$, 
\begin{equation}
[\Delta F]^2 = (F^2)-(F)^2~~~.
\label{equationb}
\end{equation}
\end{mathletters}
We shall use two other parameterizations for the state $|z\rangle$ in what 
follows. Since $(F)$ is homogeneous of degree zero in both 
$z^{\alpha}$ and $\overline z^{\alpha}$, let us define new 
complex coordinates $t^j$ by  
\begin{equation}t^j=z^j/z^0,~~ \overline t^j=\overline z^j 
/ \overline z^0~,~~~j=1,...,n. ~~~
\end{equation}
Next, it is convenient to 
split each  of the complex numbers $t^j$ into its real and imaginary 
part $t^j_R,~t^j_I$, and to introduce a $2n$ component real vector 
$x^a,~a=1,...,2n$ defined by $x^1=t^1_R,~x^2=t^1_I,~x^3=t^2_R,~
x^4=t^2_I,...,x^{2n-1}=t^n_R,~x^{2n}=t^n_I$.   Clearly, specifying 
the projective coordinates $t^j$ or $x^a$ uniquely determines the 
unit ray containing the unnormalized state $|z\rangle$, while leaving 
the normalization and ray representative of the state $|z\rangle$ 
unspecified.   

As discussed in Refs. [2], projective Hilbert space is also a Riemannian  
space with respect to the Fubini-Study metric $g_{\alpha \beta}$, defined 
by the line element 
\begin{mathletters}
\label{allequations}
\begin{equation}
ds^2= g_{\alpha \beta} d\overline z^{\alpha} dz^{\beta}
\equiv 4\left( 1- { | \langle z | z+dz \rangle |^2 \over \langle z |z \rangle 
\langle z+dz | z+dz \rangle } \right) ~~~.
\label{equationa}
\end{equation}
Abbreviating $\overline z^{\gamma} z^{\gamma} \equiv  \overline z \cdot z$, 
a simple calculation gives 
\begin{equation}
g_{\alpha \beta}=4(\delta_{\alpha \beta} \overline z \cdot z
-z^{\alpha} \overline z^{\beta})/(\overline z \cdot z)^2
=4 {\partial \over \partial \overline z^{\alpha} }
{\partial \over \partial z^{\beta} } \log \overline z \cdot z~~~.
\label{equationb}
\end{equation}
\end{mathletters}
Because of the homogeneity conditions $\overline z^{\alpha} g_{\alpha \beta} 
=z^{\beta} g_{\alpha \beta}=0$, the metric $g_{\alpha \beta}$ is not 
invertible, but if we hold the coordinates $\overline z^0,~z^0$ fixed in  
the variation of Eq.~(3a) and go over to the 
projective coordinates $t^j$, we can rewrite the line element of Eq.~(3a) 
as 
\begin{mathletters}
\label{allequations}
\begin{equation}
ds^2=g_{jk}d\overline t^j dt^k~~~,
\label{equationa}
\end{equation}
with the invertible metric [5] 
\begin{equation}
g_{jk}={4[(1+\overline t^{\ell} t^{\ell}) \delta_{jk} - t^j \overline t^k ]
\over (1+\overline t^m t^m)^2 }~~~,
\label{equationb}
\end{equation}
with inverse 
\begin{equation}
g^{jk}={1 \over 4} (1+\overline t^m t^m) (\delta_{jk} + t^j \overline t^k)
~~~.
\label{equationc}
\end{equation}
Reexpressing the complex projective coordinates $t^j$ in terms of the 
real coordinates $x^a$, the line element can be rewritten as 
\begin{eqnarray}
ds^2=&&g_{ab}dx^adx^b~~~,\nonumber\\
g_{ab}=&&{4[(1+x^dx^d)\delta_{ab}-(x^ax^b+\omega_{ac}x^c\omega_{bd}x^d)] 
\over (1+x^e x^e)^2}~~~,\nonumber\\
g^{ab}=&&{1\over 4} (1+x^e x^e)(\delta_{ab}+
x^ax^b+\omega_{ac}x^c\omega_{bd}x^d)~~~.
\label{equationd}
\end{eqnarray}
\end{mathletters}
Here $\omega_{ab}$ is a numerical tensor whose only nonvanishing elements are  
$\omega_{a=2j-1 ~b=2j}=1$ and $\omega_{a=2j~b=2j-1}=-1$
for $j=1,...,n$.  As discussed 
by Hughston, one can define a complex structure $J_a^{~b}$ over the entire 
projective Hilbert space for which $J_a^{~c}J_b^{~d}g_{cd}=g_{ab},$   
$J_a^{~b}J_b^{~c}=-\delta_a^c$, 
such that $\Omega_{ab}=g_{bc} J_a^{~c}$ and 
$\Omega^{ab}=g^{ac}J_c^{~b}$ are antisymmetric tensors.  At $x=0$, the metric 
and complex structure take the values 
\begin{eqnarray}
g_{ab}=&&4 \delta_{ab}~,~~g^{ab}={1 \over 4} \delta_{ab}~~~,\nonumber\\
J_a^{~b}=&&\omega_{ab}~,~~\Omega_{ab}=4\omega_{ab}~,
~~\Omega^{ab}={1\over 4}\omega_{ab}~~~.
\end{eqnarray}

Returning to Eq.~(1a), we shall now derive some identities that are 
central to what follows.  Differentiating Eq.~(1a) with respect to 
$\overline z^{\alpha}$, with respect to $z^{\beta}$, and with respect to both 
$\overline z^{\alpha}$ and $z^{\beta}$, we get 
\begin{mathletters}
\label{allequations}
\begin{eqnarray}
\langle z | z \rangle {\partial (F) \over \partial \overline z^{\alpha}}
=&&F_{\alpha \beta} z^{\beta} - (F) z^{\alpha}~~~,\nonumber\\
\langle z | z \rangle {\partial (F) \over \partial  z^{\beta}}=&&
\overline z^{\alpha} F_{\alpha \beta}  - (F) \overline z^{\beta}~~~,\nonumber\\
\langle z | z \rangle^2 {\partial^2 (F) \over \partial \overline z^{\alpha} 
\partial z^{\beta} }=&&
\langle z |z \rangle [F_{\alpha \beta}-\delta_{\alpha \beta} (F) ]
+2z^{\alpha} \overline z^{\beta} (F) - \overline z^{\gamma} F_{\gamma \beta} 
z^{\alpha}-\overline z^{\beta} F_{\alpha \gamma} z^{\gamma}~~~.
\label{equationa}
\end{eqnarray}
Writing similar expressions for a second operator expectation $(G)$, 
contracting in various combinations with the relations of Eq.~(6a), and 
using the homogeneity conditions 
\begin{equation}
\overline z^{\alpha} {\partial (F) \over \partial \overline z^{\alpha} }
=z^{\beta} {\partial (F) \over \partial z^{\beta} }
=\overline z^{\alpha} {\partial^2 (F) \over \partial \overline z^{\alpha} 
\partial z^{\beta}}
=z^{\beta} {\partial^2 (F) \over  \partial \overline z^{\alpha} 
\partial z^{\beta} } =0
\label{equationb}
\end{equation}
\end{mathletters}
to eliminate derivatives with respect to $\overline z^0,~z^0$, we get 
the following identities,
\begin{mathletters}
\label{allequations}
\begin{eqnarray}
-i(FG-GF)&&=-i \langle z| z \rangle \left( 
{\partial (F) \over \partial z^{\alpha}} 
{\partial (G) \over \partial \overline z^{\alpha}} - 
{\partial (G) \over \partial z^{\alpha}} 
{\partial (F) \over \partial \overline z^{\alpha}} \right) 
=2\Omega^{aAb} \nabla_a (F) \nabla_b (G)~~~,\nonumber\\
(FG+GF)-2(F)(G)&&= \langle z| z \rangle \left( 
{\partial (F) \over \partial z^{\alpha}} 
{\partial (G) \over \partial \overline z^{\alpha}} + 
{\partial (G) \over \partial z^{\alpha}} 
{\partial (F) \over \partial \overline z^{\alpha}} \right) 
=2g^{ab} \nabla_a (F) \nabla_b (G)~~~,\nonumber\\
(FGF)-(F^2)(G)&&-(F)(FG+GF)+2(F)^2(G)\nonumber\\
&&=\langle z | z \rangle ^2 
{\partial (F) \over \partial z^{\alpha}}
{\partial^2 (G) \over \partial \overline z^{\alpha} \partial z^{\beta}}
{\partial (F) \over \partial \overline z^{\beta}}
=2\nabla^a (F) \nabla^b (F) \nabla_a \nabla_b (G),
\label{equationa}
\end{eqnarray}
with $\nabla_a$ the covariant derivative with respect to the Fubini-Study 
metric.  It is not necessary to use the detailed form of the 
affine connection 
to verify the right hand equalities in these identities, because since $(G)$ 
is a Riemannian scalar, $\nabla_a \nabla_b (G)$$ =\nabla_a \partial_b (G)$, 
and since projective Hilbert space is a homogeneous manifold, it suffices 
to verify the identities at the single point $x=0$, where the affine 
connection vanishes and thus 
$\nabla_a \nabla_b (G)=\partial_a \partial_b (G)$.
Using Eqs.~(7a) and the chain rule we also find 
\begin{equation}
-\nabla^a [(F^2)-(F)^2] \nabla_a (G)=
-{1\over 2} (F^2 G+G F^2) +(F^2)(G) + (F) (FG+GF) -2(F)^2(G)~~~,
\label{equationb}
\end{equation}
which when combined with the final identity in Eq.~(7a) gives 
\begin{equation}
\nabla^a(F) \nabla^b(F) \nabla_a \nabla_b (G)
-{1\over 2} \nabla^a [(F^2)-(F)^2] \nabla_a (G)= 
-{1\over 4}([F,[F,G]])~~~,
\label{equationc}
\end{equation}
\end{mathletters}
the right hand side of which vanishes when the operators $F$ and $G$ 
commute [6].  

Let us now turn to Hughston's stochastic differential equation, which 
in our notation is 
\begin{mathletters}
\label{allequations}
\begin{equation}
dx^a=[2 \Omega^{ab}\nabla_b(H)-{1\over 4}\sigma^2 \nabla^aV]dt
+\sigma\nabla^a(H) dW_t~~~,
\label{equationa}
\end{equation}
with $W_t$ a Brownian motion or Wiener process, with $\sigma$ a parameter 
governing the strength of the stochastic terms, with $H$ the Hamiltonian 
operator and $(H)$ its expectation, and with  $V$ the 
variance of the Hamiltonian, 
\begin{equation}
V=[\Delta H]^2=(H^2)-(H)^2~~~.
\label{equationb}
\end{equation}
\end{mathletters}
When the parameter $\sigma$ is zero, Eq.~(8a) is just the transcription 
of the Schr\"odinger equation to projective 
Hilbert space.  For the time evolution of a 
general function $G[x]$, we get by 
Taylor expanding $G[x+dx]$ and using the It\^o stochastic calculus rules 
\begin{mathletters}
\label{allequations}
\begin{equation}
[dW_t]^2=dt~,~~[dt]^2=dtdW_t=0~~~,
\label{equationa}
\end{equation}
the corresponding stochastic differential equation
\begin{equation}
dG[x]=\mu dt + \sigma \nabla_aG[x]\nabla^a(H) dW_t~~~,
\label{equationb}
\end{equation}
with the drift term $\mu$ given by 
\begin{equation}
\mu=2 \Omega^{ab} \nabla_aG[x]\nabla_b(H)-{1\over 4}
\sigma^2\nabla^aV\nabla_a
G[x]+{1\over 2}\sigma^2 \nabla^a(H)\nabla^b(H)\nabla_a\nabla_bG[x]~~~.
\label{equationc}
\end{equation}
\end{mathletters}
Hughston shows that with the $\sigma^2$ part of the drift term 
chosen as in Eq.~(8a), the drift term $\mu$ in Eq.~(9c) vanishes for the 
special case $G[x]=(H)$, guaranteeing conservation 
of the expectation of the energy with respect to the stochastic evolution 
of Eq.~(8a).  But referring to Eq. (7c) and the first identity in Eq.~(7a),  
we see that in fact 
a much stronger result is also true, namely that $\mu$ vanishes [and thus
the stochastic process of Eq.~(9b) is a martingale] whenever 
$G[x]=(G)$, with $G$ any operator that commutes with the Hamiltonian $H$.   

Let us now make two applications of this fact.  First, taking $G[x]=V=
(H^2)-(H)^2$, we see that the contribution from $(H^2)$ to $\mu$ 
vanishes, so the drift term comes entirely from $-(H)^2$.  
Substituting this into $\mu$ gives $-2(H)$ times the drift term produced by 
$(H)$, which is again zero, plus an extra term 
\begin{mathletters}
\label{allequations}
\begin{equation}
-\sigma^2 \nabla^a(H)\nabla ^b(H)\nabla_a(H)\nabla_b(H)
=-\sigma^2V^2~~~,
\label{equationa}
\end{equation}
where we have used the relation $V=\nabla_a(H)\nabla^a(H)$ which follows 
from the $F=G=H$ case of the middle identity of Eq.~(7a).  Thus the 
variance $V$ of the Hamiltonian satisfies the stochastic differential 
equation, derived by Hughston by a more complicated method, 
\begin{equation}
dV=-\sigma^2 V^2 dt + \sigma \nabla_aV\nabla^a(H) dW_t~~~.
\label{equationb}
\end{equation}
This implies that the expectation $E[V]$ with respect to the stochastic 
process obeys 
\begin{equation}
E[V_t]=E[V_0]-\sigma^2 \int_0^t ds E[V_s^2]~~~,
\label{equationc}
\end{equation}
which using the inequality $0\leq E[\{V-E[V]\}^2]=E[V^2]-E[V]^2$ 
gives the inequality 
\begin{equation}
E[V_t] \leq E[V_0] -\sigma^2 \int_0^t ds E[V_s]^2~~~.
\label{equationd}
\end{equation}
\end{mathletters}
Since $V$ is necessarily positive, Eq.~(10d) implies that $E[V_{\infty}]=0$, 
and again using positivity of $V$ this implies that $V_s$ 
vanishes as $s \to \infty$, apart from a set of outcomes 
of probability measure zero.  Thus, as concluded by 
Hughston, the stochastic term in his equation drives the system, as $t \to 
\infty$, to an energy eigenstate.  

As our second application of the vanishing of the drift term $\mu$ for 
expectations of operators that commute with $H$, let us consider the 
projectors $\Pi_e\equiv |e\rangle \langle e| $ on a complete set of 
energy eigenstates $|e \rangle$.  By definition, these projectors all 
commute with H, and so the drift term $\mu$ vanishes in the stochastic 
differential equation for $G[x]=(\Pi_e)$, and consequently the expectations   
$E[(\Pi_e)]$ are time independent; additionally, by completeness of   
the states $|e\rangle$, we have $\sum_e (\Pi_e)=1$.  
But these are just the conditions for   
Pearle's [4] gambler's ruin argument to apply.  At time zero, 
$E[(\Pi_e)]=(\Pi_e)\equiv p_e$
is the absolute value squared of the quantum mechanical amplitude  
to find the initial state in energy eigenstate $|e \rangle$.  At $t=\infty$, 
the system always evolves to an energy eigenstate, with the eigenstate  
$|f\rangle $ occurring with some probability $P_f$.  The expectation 
$E[(\Pi_e)]$, evaluated at infinite time, is then  
\begin{equation}
E[(\Pi_e)]=1 \times P_e + \sum_{f \neq e} 0 \times P_f = P_e~~~;
\end{equation}
hence $p_e=P_e$ for each  $e$ and the state collapses into energy eigenstates 
at $t=\infty$ with probabilities given by the usual quantum mechanical 
rule applied to the initial wave function [7].  

Let us now examine the structure of Hughston's 
equation for a Hilbert space constructed as the direct product of  
independent subsystem Hilbert spaces, so that 
\begin{mathletters}
\label{allequations}
\begin{eqnarray}
|z\rangle =&& \prod_{\ell} |z_{\ell} \rangle~~~,\nonumber\\
H=&&\sum_{\ell} H_{\ell}~~~,
\label{equationa}
\end{eqnarray}
with $H_{\ell}$ acting as the unit operator on the states $|z_{k}\rangle ~,~~
k \neq \ell$.  Then a simple calculation shows that the expectation 
of the Hamiltonian $(H)$ and its variance $V$  are both 
additive over the subsystem Hilbert spaces, 
\begin{eqnarray}
(H)=&&\sum_{\ell} (H_{\ell})_{\ell}~~~,\nonumber\\
V=\sum_{\ell} V_{\ell} =&&\sum_{\ell}[ (H_{\ell}^2)_{\ell} 
-(H_{\ell})_{\ell}^2]~~~,
\label{equationb}
\end{eqnarray}
\end{mathletters}
with $(F_{\ell})_{\ell}$ the expectation of the operator $F_{\ell}$ 
formed according to Eq.~(1a) with respect 
to the subsystem wave function $|z_{\ell}\rangle$.  In addition, 
the Fubini-Study line element is also additive over the subsystem Hilbert 
spaces, since [8]
\begin{eqnarray}
1-ds^2/4=&& {| \langle z | z+dz \rangle |^2 \over \langle z |z \rangle 
\langle z+dz | z+dz \rangle } =\prod_{\ell}  
{ | \langle z_{\ell} | z_{\ell}+dz_{\ell} \rangle |^2 \over 
\langle z_{\ell} |z_{\ell} \rangle \langle z_{\ell}+dz_{\ell} 
| z_{\ell}+dz_{\ell} \rangle }\nonumber\\
=&&\prod_{\ell}[1-ds_{\ell}^2/4]=1-[\sum_{\ell} ds_{\ell}^2]/4 +{\rm O}(ds^4)
~~~.
\end{eqnarray}
As a result of Eq.~(13), the metric $g^{ab}$ and complex  
structure $\Omega^{ab}$ block diagonalize over the independent subsystem 
subspaces. Equation (12b) then implies that Hughston's stochastic extension  
of the Schr\"odinger equation given in Eq.~(8a) 
separates into similar equations for the subsystems, that do not refer 
to one another's $x^a$ coordinates, but are correlated only through the  
common Wiener process $dW_t$ that appears in all of them.  Under the  
assumption [9] that $\sigma \sim M_{\rm Planck}^{-1/2}$ in microscopic 
units with $\hbar =c=1$, these correlations will be very small; it will be 
important to analyze whether they can have observable physical 
consequences on laboratory or cosmological scales [10].  

To summarize, we have shown that Hughston's stochastic extension of the 
Schr\"odinger equations has properties that make it a viable physical model 
for state vector reduction.   This opens the challenge of seeing whether 
it can be derived as a phenomenological approximation to a fundamental 
pre-quantum dynamics.  Specifically, we suggest that since 
Adler and Millard [11] have argued that quantum mechanics can emerge as 
the thermodynamics of an underlying non-commutative operator dynamics, 
it may be possible to show that Hughston's stochastic process is the  
leading statistical fluctuation correction to this thermodynamics.  

\acknowledgments
This work was supported in part by the Department of Energy under
Grant \#DE--FG02--90ER40542.  One of us (S.L.A.) wishes to thank J. 
Anandan for conversations introducing him to the Fubini-Study metric.  
The other (L.P.H.) wishes to thank P. Leifer for many discussions on 
the properties of the complex projective space.


\begin{references}
\bibitem{[1]}  For a representative, but not exhaustive, survey of the earlier 
literature, see the papers of Di\'osi, Ghirardi et. al., Gisin, 
Pearle, and Percival cited by Hughston, Ref. [3] below.
\bibitem{[6]}  An alternative demonstration of this result uses the
fact,  noted by 
Hughston [3], that $\xi^a_F\equiv \Omega^{ab} \nabla_b(F)$ is a 
Killing vector obeying $\nabla_c\xi_F^a+\nabla^a \xi_{Fc}=0$.  First rewrite 
$\nabla^a(F)\nabla^b(F)\nabla_a\nabla_b(G)$ as $\Omega^{ca} \Omega^{eb} 
\xi_{Fc}\xi_{Fe} \nabla_a\nabla_b(G)$ $=\xi_{Fc}\xi_{Fe}\Omega^{ca} 
\nabla_a \xi_G^e$.  By the Killing vector property, this becomes 
$-\xi_{Fc}\xi_{Fe}\Omega^{ca}\nabla^e \xi_{Ga}$, which can be rewritten 
as $-\nabla^e[\xi_{Fe}\xi_{Fc}\Omega^{ca}\xi_{Ga}]$
$+\nabla^e\xi_{Fe}\, \xi_{Fc}\Omega^{ca}\xi_{Ga}$
$+\xi_{Fe}\nabla^e\xi_{Fc}\,\Omega^{ca}\xi_{Ga}$.   
When $F$ and $G$ commute,  
the first two terms vanish by the first identity in Eq.~(7a), while using 
the Killing vector property for $\xi_F$ in the third term gives 
$-\nabla_c\xi_F^e\, \xi_{Fe} \Omega^{ca} \xi_{Ga}$ = 
$-(1/2)\nabla_c[\xi_F^e \xi_{Fe}]\Omega^{ca}J_a^{~b} \nabla_b(G)$, 
which using $\Delta F=\xi_F^e\xi_{Fe}$ reduces to $(1/2)\nabla_c[\Delta F] 
\nabla^c (G)$.\hfill\break
\bibitem{[7]} This conclusion readily generalizes to the stochastic equation 
$dx^a=[2 \Omega^{ab}\nabla_b(H)-{1\over 4}\sigma^2 \sum_j \nabla^aV_j]dt
+\sigma\sum_j \nabla^a(H_j) dW_t^j~~~,$ with the $H_j$ a set of
mutually commuting operators that commute with H, with 
$V_j =(H_j^2)-(H_j)^2$, and with the $dW_t^j$ independent Wiener processes
obeying $dW_t^j dW_t^k=\delta^{jk}dt$~.
\bibitem{[8]} An alternative way to see this is to use the identity 
$\log \overline z \cdot z =\log \prod_{\ell} \overline z_{\ell} 
\cdot z_{\ell} =$ $ \sum_{\ell} \log \overline z_{\ell}\cdot z_{\ell}$ 
in Eq.~(3b), along with a change of 
variable from $z$ to the $z_{\ell}$'s.\hfill\break
\end{references}
\end{document}